\documentclass[9pt,twocolumn,twoside]{opticajnl}
\journal{josab} 

\setboolean{shortarticle}{false}

\usepackage{lineno}

\usepackage{epsf,graphicx,epstopdf,xcolor}

\renewcommand{\vec}[1]{{\bf #1}}
\renewcommand{\d}{{\rm d}}
\renewcommand{\i}{{\rm i}}
\newcommand{\pard}{{\rm\partial}}
\newcommand{\e}{\mathop{\rm e}\nolimits}

\title{Interaction of light with subwavelength particles: Revealing the physics of the electric dipole moment in the classical scattering problem}

\author{Yuriy~A.~Akimov}

\affil[1]{Institute of High Performance Computing (IHPC), Agency for Science, Technology and Research (A*STAR),
1 Fusionopolis Way, \#16-16 Connexis, Singapore 138632}

\affil[*]{yuriy.a.akimov@gmail.com}

\begin{abstract}
Scattering problems are the classical tools for modeling of light-matter interaction. 
In this paper, we investigate the solution of the dipole scattering problem under different incident radiations. 
In particular, we compare the two cases of incident plane and spherically incoming fields. 
With this comparison, we disclose the two distinct groups of current-sourced and current-free scattered fields, which exhibit independent dynamics and dissimilar effects of the scatterer. 
We demonstrate how these fields by interfering each other make the resultant electric dipole moment of the scattered fields resonant and, thus, give rise to all the spectral features observed in the classical solution for dipole scattering of light.       
\end{abstract}

\setboolean{displaycopyright}{false} 
\doi{https://doi.org/10.1364/JOSAB.524989}

\begin{document}

\maketitle

\section{Introduction}

Classical scattering problems play a key part in our understanding of light-matter interaction. 
Scattering by subwavelength particles explains the blue color of the daytime sky and the reddish color of the low Sun, as well as enables the qualitative description of matter polarization at the atomic scale \cite{Landau:1984, Jackson:1999, Bohren:1998}.
Despite the wide use of scattering problems, their solutions still keep a number of unanswered fundamental questions. 
For instance, subwavelength particles are known to exhibit frequency-selective effects such as strong extinction \cite{Bohren:1998, Bohren:1983} or nonradiating anapole states \cite{Luk'yanchuk:2017, Miroshnichenko:2015}.
These effects are commonly explained by the interference of the particle's multipole currents induced by the incident fields \cite{Jackson:1999, Grahn:2012}.
However, if we consider the size-dependence of multipole current moments, we can notice a contradiction in this explanation. 
For a sphere of radius $R$ smaller than the wavelength of incident light, the lowest electric moment, the dipolar one, is proportional to $R^3$ \cite{Luk'yanchuk:2017, Miroshnichenko:2015, Grahn:2012}.
The higher-order electric moments have stronger dependence. 
For instance, the next-to-dipolar moment, the toroidal one, is proportional to $R^5$ \cite{Luk'yanchuk:2017, Miroshnichenko:2015, Grahn:2012}. 
Thus, particles with small $R$ naturally possess dominant electric dipole and negligibly small higher-order currents. 
From the one side, this size-dependence makes any pronounced interference of current moments forbidden for deeply subwavelength particles with $R \rightarrow 0$. 
From the other side, resonant effects of deeply subwavelength metal nanoparticles with dominant electric dipole moments are highly pronounced and include both nonradiating and super-radiating states \cite{Tribelsky:2006, Luk'yanchuk:2007, Akimov:2012, Kolwas:2013}.
This suggests that the resonances observed in classical scattering problems are caused by processes different from the interference of current multipoles.
Those processes are expected to bring resonances to the level of a single electric dipole moment without involving any higher-order currents.    
To address this issue, we reconsider the classical quasi-static solution of dipole scattering by including the radiative decay to the scattered fields that allows us to look at the solution from a new physically wider perspective. 
Finally, this help us clarify the failure of the current moment decomposition for characterization of the dipolar fields obtained in the classical problem of light scattering.  

\section{Dipole scattering}

Let us consider the quasi-static regime of scattering \cite{Landau:1984,Bohren:1998,Jackson:1999} given by Maxwell's equations for a spherical particle of radius $R$ and dielectric permittivity $\varepsilon_i$ embedded in a material of dielectric permittivity $\varepsilon_e$. If the particle is exposed to a harmonic incident electric field $\vec E_{\rm inc}$, then the total electric field outside the particle, $\vec E(r>R)=\vec E_{\rm ext}$, is given by the superposition of the incident $\vec E_{\rm inc}$ and scattered $\vec E_{\rm sca}$ fields: 
\begin{equation}
	\vec E_{\rm ext}=\vec E_{\rm inc}+\vec E_{\rm sca}.
\end{equation}
For deeply subwavelength particles with $k_e R\ll 1$, where $k_e=2\pi\sqrt{\varepsilon_e}/\lambda$ is the external medium wavenumber at the vacuum wavelength $\lambda$, the scattered field is dominated by the quasi-static dipolar fields \cite{Landau:1984,Bohren:1998,Jackson:1999}. At small distances $r$ compared to the wavelength $\lambda$, when $k_er\ll1$ holds, the scattered field is given by 
\begin{eqnarray}
	\vec E_{\rm sca}=AE_0\frac{R^3}{r^3}[\vec e_r(2+\i f_{\rm rad}(r))\cos\theta+\hspace{1cm}\nonumber\\
	\vspace{3cm}\vec e_\theta(1-\i f_{\rm rad}(r))\sin\theta]\e^{-\i\omega t},\label{Esca}
\end{eqnarray}
where $A$ is the scattered field amplitude (see Appendices \ref{AppA} and \ref{AppB} for derivation) yet to be determined, $E_0$ is the amplitude of the incident field assumed to be $z$-polarized, $\omega=2\pi c/\lambda$ is the angular frequency of the incident field with $c$ being the speed of light, $(r,\theta,\phi)$ are the spherical coordinates, and $\vec e_r, \vec e_\theta, \vec e_\phi$ are the respective spherical unit vectors. 
In addition to the dipolar decay, $\vec E_{\rm sca}\propto r^{-3}$, prevailing in the scattered field, we also account for the radiative decay given by small imaginary contributions with $f_{\rm rad}(r)=(2/3)(k_e r)^3\ll1$ being the leading radiative terms in the full-wave solution of dipolar fields, as shown in Appendix \ref{AppB}.
The dominance of the dipolar fields for small spheres holds for the electric field excited inside the particle, $\vec E(r< R)=\vec E_{\rm int}$, if $|k_i| R\ll 1$ with $k_i=2\pi\sqrt{\varepsilon_i}/\lambda$. In this case, $\vec E_{\rm int}$ is given by the uniform $z$-polarized field: 
\begin{equation}
	\vec E_{\rm int}=BE_0(\vec e_r \cos\theta-\vec e_\theta \sin\theta)\e^{-\i\omega t},\label{Eint}
\end{equation}
where $B$ is the internal field amplitude (see Appendix \ref{AppB} for details). The unknown amplitudes $A$ and $B$ in Eqs.~(\ref{Esca}) and (\ref{Eint}) can be determined by applying the boundary conditions consisting of the continuity of (i) the normal component $\vec e_r\cdot\varepsilon_0\varepsilon\vec E$ of the electric displacement field, where the space-dependent permittivity $\varepsilon$ is given by $\varepsilon_i$ at $r< R$ and $\varepsilon_e$ at $r> R$, and (ii) the tangential components $\vec e_\theta\cdot \vec E$ and $\vec e_\phi\cdot \vec E$ of the electric field at the sphere interface $r=R$.

As the fields scattered by small particles are dipolar by their composition, they can be considered as the fields emitted by a point electric dipole (see, e.g., Section 5.2 in Ref.~\cite{Bohren:1998} or Section 4.5 in Ref.~\cite{Jackson:1999}) with the effective dipole moment  
\begin{equation}
	\vec p_{\rm eff}=\vec e_z 4\pi\varepsilon_0\varepsilon_e A R^3E_0 \e^{-\i\omega t}.
\end{equation}
Being the scattered-field-based electric dipole moment, $\vec p_{\rm eff}$ provides complete description of the processes underlying light scattering by small spheres \cite{Bohren:1998,Jackson:1999}. 

\section{Scattering current approach}

An alternative to the effective dipole description is the scattering current approach \cite{Jackson:1999, Grahn:2012}.
If the scattered field $\vec E_{\rm sca}$ is defined in the entire space, i.e. when $\vec E=\vec E_{\rm inc}+\vec E_{\rm sca}$ for both the internal and external domains, then Maxwell's equations
\begin{eqnarray}
	&\displaystyle 
	\nabla\times\vec H=-\i\omega\varepsilon_0\varepsilon\vec E,\quad
	\nabla \times\vec E=\i\omega\mu_0\vec H,\label{Maxwell}
\end{eqnarray}
where $\vec E$ and $\vec H$ are the total electric and magnetic fields excited at angular frequency $\omega$,
can be split into
\begin{eqnarray}
	&\nabla\times \nabla\times\vec E_{\rm inc}-\omega^2\mu_0\varepsilon_0\varepsilon_e\vec E_{\rm inc}=0,\label{Maxwell0}\\
	&\nabla\times \nabla\times\vec E_{\rm sca}-\omega^2\mu_0\varepsilon_0\varepsilon_e\vec E_{\rm sca}=\i\omega\mu_0\vec J_{\rm sca},\label{Maxwell1}
\end{eqnarray}
with the current density 
\begin{eqnarray}
	\vec J_{\rm sca}=-\i\omega\varepsilon_0(\varepsilon-\varepsilon_e)\vec E,\label{J_sca}
\end{eqnarray}
serving as the source of the scattered fields, where $\varepsilon_0$ and $\mu_0$ are the electric and magnetic constants.
Noteworthy that the scattering current density $\vec J_{\rm sca}$ appears nonzero inside the sphere only, where the total field $\vec E$ is given by the uniform $\vec E_{\rm int}$.
If we decompose the current density $\vec J_{\rm sca}$ over different-order moments, the latter can be used for description of the scattered fields and optical effects associated with them. 
For deeply subwavelength spheres with $k_e R,|k_i| R\ll 1$, the scattering current is dominated by the electric dipole moment
\begin{eqnarray}
	&\displaystyle
	\vec p_{\rm sca}=
	\frac{\i}{\omega}\int \vec J_{\rm sca} \;\d^3 r=
	\vec e_z \frac{4\pi}{3} \varepsilon_0(\varepsilon_i-\varepsilon_e)B R^3 E_0\e^{-\i\omega t}.
\end{eqnarray}
The moment $\vec p_{\rm sca}$ is widely used alongside with $\vec p_{\rm eff}$ as a measure of the dipolar fields scattered by small particles, assuming that the two moments coincide \cite{Landau:1984,Bohren:1998,Jackson:1999}. 

\section{Scattering of plane fields}

Although the scattering current approach has gotten wide acceptance in the optics community, its hypothesis of equal scattering and effective dipole moments remains unproven. To test it, we can consider dipole scattering of plane incident fields (see Appendix \ref{AppB} for details):
\begin{equation}
	\vec E^{\rm pl}_{\rm inc}=E_0(\vec e_r \cos\theta-\vec e_\theta \sin\theta)\e^{-\i\omega t},\label{plane incident}
\end{equation}
by deeply subwavelength spheres. Application of the boundary conditions for electric fields at the sphere interface gives us the following amplitudes of the scattered and internal fields:
\begin{eqnarray}
	A^{\rm pl}=A(\vec E_{\rm inc}=\vec E^{\rm pl}_{\rm inc})=\frac{A_{\rm qs}}{1-\i A_{\rm qs} f_{\rm rad}(R)},\\
	B^{\rm pl}=B(\vec E_{\rm inc}=\vec E^{\rm pl}_{\rm inc})=\frac{B_{\rm qs}}{1-\i A_{\rm qs} f_{\rm rad}(R)},
\end{eqnarray}
where 
\begin{eqnarray}
	A_{\rm qs}=\frac{\varepsilon_i-\varepsilon_e}{\varepsilon_i+2\varepsilon_e},\quad
	B_{\rm qs}=\frac{3\varepsilon_e}{\varepsilon_i+2\varepsilon_e},\label{AB_qs}
\end{eqnarray}
are the quasi-static amplitudes commonly obtained in the limiting case of $f_{\rm rad}(R)\rightarrow 0$ \cite{Landau:1984,Bohren:1998,Jackson:1999}. 

The scattered fields and scattering currents derived result in the equal effective and scattering electric dipole moments: 
\begin{equation}
	\vec p^{\rm pl}_{\rm eff}=\vec p^{\rm pl}_{\rm sca}=\vec e_z\frac{4\pi\varepsilon_0\varepsilon_e R^3 E_0A_{\rm qs}}{1-\i A_{\rm qs} f_{\rm rad}(R)}\e^{-\i\omega t}.\label{p_sca_pl}
\end{equation}
The coincidence of $\vec p^{\rm pl}_{\rm eff}$ and $\vec p^{\rm pl}_{\rm sca}$ is commonly interpreted as the validation of the scattering current approach \cite{Bohren:1998}. Eventually, this approach is widely used, e.g., in discrete dipole approximation for simulation of scattering by large objects of arbitrary shape \cite{Draine:1988,Draine:1994,Zubko:2010,Evlyukhin:2011}, in multipole current decomposition of scattering peculiarities, such as nonradiating anapole configurations, bright and dark mode resonances, etc. \cite{Evlyukhin:2016,Fernandez-Corbaton:2017,Evlyukhin:2019,Liu:2020,Alaee:2020,Zenin:2020,Gurvitz:2019,Basharin:2023,Ospanova:2023}. 

\begin{figure*}[t]
	\centering\includegraphics[width=0.85\textwidth,clip,trim={0cm 10.8cm 0cm 0cm}]{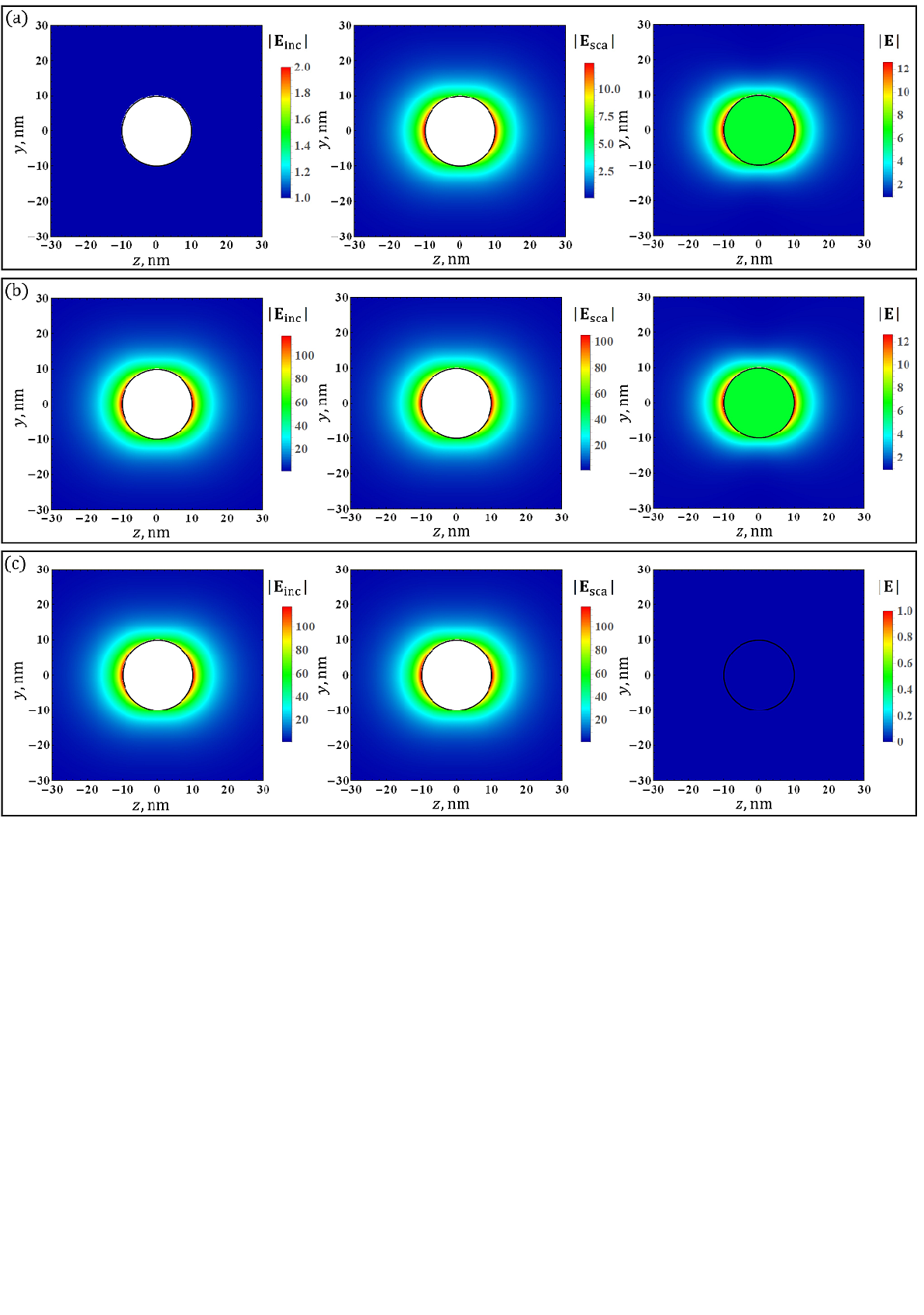}
	\caption{Distribution of dipolar electric fields at $\lambda=400$ nm given by the quasi-static solution for (a) planar and (b) spherically incoming fields incident on a silver particle of $R=10$ nm embedded in SiO$_2$. 
(c) Difference of the fields plotted in (a) and (b).     
	\label{Fig1}}
\end{figure*}

\section{Scattering of spherically incoming fields}

Despite the demonstrated $\vec p^{\rm pl}_{\rm eff}=\vec p^{\rm pl}_{\rm sca}$, the general assumption of the scattering current approach, $\vec p_{\rm eff}=\vec p_{\rm sca}$, for an {\it arbitrary} incident field remains an unproven hypothesis. 
To prove it, Eq.~(\ref{Maxwell1}) must be solved in a general form for an arbitrary incident field. 
Since such a solution is not as straightforward, the correlation between $\vec p_{\rm eff}$ and $\vec p_{\rm sca}$ remains under question and may potentially not hold for non-plane incident fields. 
To further test it, we change the incident plane fields to the spherically incoming dipolar ones (see Appendices \ref{AppB} for details): 
\begin{eqnarray}
	\vec E^{\rm sph,in}_{\rm inc}=\frac{\i E_0}{2f_{\rm rad}(r)}[\vec e_r(2-\i f_{\rm rad}(r))\cos\theta+\hspace{1cm}\nonumber\\
	\vec e_\theta(1+\i f_{\rm rad}(r))\sin\theta]\e^{-\i\omega t},\label{spherical in}
\end{eqnarray}
that similarly result in dipolar scattered fields and can be characterized with $\vec p_{\rm eff}$ and $\vec p_{\rm sca}$. 
In this case, we obtain the following amplitudes of the scattered and internal fields: 
\begin{eqnarray}
	&\displaystyle
	A^{\rm sph,in}=A(\vec E_{\rm inc}=\vec E^{\rm sph,in}_{\rm inc})=
	\frac{1+\i A_{\rm qs} f_{\rm rad}(R)}{2\i f_{\rm rad}(R)[1-\i A_{\rm qs} f_{\rm rad}(R)]},~~~\\
	&\displaystyle
	B^{\rm sph,in}= B(\vec E_{\rm inc}=\vec E^{\rm sph,in}_{\rm inc})=\frac{B_{\rm qs}}{1-\i A_{\rm qs} f_{\rm rad}(R)}.
\end{eqnarray}
The scattered fields result in the effective electric dipole moment
\begin{eqnarray}
\vec p^{\rm sph,in}_{\rm eff}=\vec e_z\frac{2\pi\varepsilon_0\varepsilon_e R^3 E_0[1+\i A_{\rm qs} f_{\rm rad}(R)]}{\i f_{\rm rad}(R)[1-\i A_{\rm qs} f_{\rm rad}(R)]}\e^{-\i\omega t},
\end{eqnarray}
while the scattering current gives us 
\begin{eqnarray}
\vec p^{\rm sph,in}_{\rm sca}=\vec e_z\frac{4\pi\varepsilon_0\varepsilon_e R^3E_0A_{\rm qs}}{1-\i A_{\rm qs} f_{\rm rad}(R)}\e^{-\i\omega t}.\label{p_sca_sphin}
\end{eqnarray} 
The observed mismatch, $\vec p^{\rm sph,in}_{\rm eff}\neq\vec p^{\rm sph,in}_{\rm sca}$, demonstrates that the scattering current approach generally fails for non-plane incident fields. 
This failure can be understood from Eq.~(\ref{J_sca}) considered from the point of view of the relation between the scattering current density $\vec J_{\rm sca}$ and the scattered field $\vec E_{\rm sca}$. 
Following Eq.~(\ref{J_sca}), the zero scattering current $\vec J_{\rm sca}$ is associated with the null {\it total} field $\vec E$, but not the null {\it scattered} field $\vec E_{\rm sca}$. 
This means that Maxwell's equation generally support existence of the scattered fields which are not current-sourced. 

The paradox of {\it source-free} scattered fields comes from the formulation of scattering problems. 
In addition to the conventional electrodynamic formulation, following which all electromagnetic fields are generated by electric currents, scattering problems introduce the so-called free fields which are sourceless. 
Namely, the incident fields used in scattering problems are given in the form of such source-free fields \cite{Landau:1984,Bohren:1998,Jackson:1999}. 
This is clearly seen in Eq.~(\ref{Maxwell0}) for the incident field, where we have zero for the source term in the right-hand size.
Eventually, scattered fields appear composed of (i) current-sourced fields that can be treated in terms of their excitation currents and (ii) source-free fields accompanied by the zero induced current. 
The latter appear in scattering problem as a part of the {\it trivial solutions} of Maxwell's equations given by the null total electric field $\vec E$ and the null scattering current $\vec J_{\rm sca}$ (see Appendices \ref{AppA} for details). 
In other words, the source-free scattered fields originate from the trivial response of Maxwell's equations to the introduced source-free incident fields.  

The trivial response of Maxwell's equations can be seen in copmarison of the fields excited by the plane and spherical dipolar incident waves. Following the obtained solutions, the total field distributions for planar and spherical dipole incident waves are exactly the same, while their scattered fields differ by the amplitude only, as shown in Fig.~\ref{Fig1}~(a) and (b) for a silver particle embedded in SiO$_2$. 
This becomes possible only if the two incident radiation differ by the field $\vec E^{\rm pl}_{\rm inc}-\vec E^{\rm sph,in}_{\rm inc}$ that causes the trivial response of Maxwell's equations with the scattered field difference $\vec E^{\rm pl}_{\rm sca}-\vec E^{\rm sph,in}_{\rm sca}$ contributed by the source-free field $-(\vec E^{\rm pl}_{\rm inc}-\vec E^{\rm sph,in}_{\rm inc})$.       
The field difference plotted in Fig.~\ref{Fig1}~(c) demonstrates the source-free contribution to $\vec E_{\rm sca}$ accompanied by the trivial contribution to $\vec E$.

\section{Scattering of spherically outgoing fields}

Now, let us separate the effects of source-free and current-sourced fields. As we discussed above, the source-free scattered fields mathematically appear as a part of the trivial solution of Maxwell's equations, where the scattered fields fully compensate a part of the incident fields. As the scattered fields are spherically outgoing, the compensated part of the incident field must be spherically outgoing as well. To demonstrate the purely source-free scattered fields, we consider scattering of the outgoing dipolar spherical field (see Appendices \ref{AppB} for details): 
\begin{eqnarray}
	\vec E^{\rm sph,out}_{\rm inc}=\frac{-\i E_0}{2f_{\rm rad}(r)}[\vec e_r(2+\i f_{\rm rad}(r))\cos\theta+\hspace{1cm}\nonumber\\
\vec e_\theta(1-\i f_{\rm rad}(r))\sin\theta]\e^{-\i\omega t}\label{spherical out}~~
\end{eqnarray}
that does not contain any spherically incoming component. 
After application of the boundary conditions, we obtain the following amplitudes of the scattered and internal fields: 
\begin{eqnarray}
	&\displaystyle
	A^{\rm sph,out}=A(\vec E_{\rm inc}=\vec E^{\rm sph,out}_{\rm inc})=\frac{\i}{2f_{\rm rad}(R)},\\
	&\displaystyle
	B^{\rm sph,out}=B(\vec E_{\rm inc}=\vec E^{\rm sph,out}_{\rm inc})=0,
\end{eqnarray}
revealing the nonzero source-free scattered field with the effective electric dipole moment
\begin{eqnarray}
\vec p^{\rm sph,out}_{\rm eff}=-\vec e_z\frac{2\pi\varepsilon_0\varepsilon_e R^3E_0}{\i f_{\rm rad}(R)}\e^{-\i\omega t}
\end{eqnarray}
and the null scattering current with the electric dipole moment
\begin{eqnarray}
\vec p^{\rm sph,out}_{\rm sca}=0.
\end{eqnarray} 
The solution obtained is trivial for electric field with the net zero total field $\vec E$ everywhere in space, as shown in Fig.~\ref{Fig1}~(c). This solution demonstrates the evident failure of the scattering current approach for description of such type of source-free scattered fields. 

The failure of the scattering current approach for description of scattered fields comes from its limitation to electrodynamic formulation where any electromagnetic fields must have sources in terms of their currents. By rejecting source-free fields, the scattering current approach naturally fails for complete description of scattering problems where free incident fields are used. To be fully compliant with the electrodynamic formulation, the scattering current approach must eliminate the use of any free fields, including the incident one. This requires Eq.~(\ref{Maxwell1}) to be rewritten for the total electric field: 
\begin{eqnarray}
	&\nabla\times \nabla\times\vec E-\omega^2\mu_0\varepsilon_0\varepsilon_e\vec E=\i\omega\mu_0\vec J_{\rm sca},\label{Maxwell2}
\end{eqnarray}
where $\vec J_{\rm sca}$ is now the source of $\vec E$, not $\vec E_{\rm sca}$. 
With this formal change, $\vec p_{\rm sca}$ appears the electric dipole moment of the {\it total} fields, not the scattered ones. 
Eventually, it clarifies the equality $\vec p^{\rm sph,in}_{\rm sca}=\vec p^{\rm pl}_{\rm sca}$ observed for the plane and spherically incoming $\vec E_{\rm inc}$, whose total fields $\vec E$ are identical, as shown in Fig.~\ref{Fig1}.

\section{Two types of scattered fields}

With the revealed composition of scattered fields, we can reconsider the solution obtained for plane incident fields. The plane incident field given by Eq.~(\ref{plane incident}) can be decomposed over dipolar spherically incoming (\ref{spherical in}) and outgoing (\ref{spherical out}) fields as follows: 
\begin{equation}
	\vec E^{\rm pl}_{\rm inc}=\vec E^{\rm sph,in}_{\rm inc}+\vec E^{\rm sph,out}_{\rm inc}.
\end{equation}
As the dipolar spherically incoming incident field does not contain any spherically outgoing component that could be compensated with the scattered fields, the part $\vec E^{\rm sph,in}_{\rm inc}$ of the plane incident field  $\vec E^{\rm pl}_{\rm inc}$ experiences purely current-sourced (electrodynamically compliant) scattering. At the same time, the spherically outgoing part $\vec E^{\rm sph,out}_{\rm inc}$ of the plane incident field is fully compensated by the scattered fields and, thus, experiences purely current-free (electrodynamically incompliant) scattering. In other words, decomposition of the plane fields into spherically incoming and outgoing fields helps us split the effective electric dipole moment of the plane incident field into two moments fundamentally different from the electrodynamics point of view:
\begin{equation}
	\vec p^{\rm pl}_{\rm eff}=\vec p^{\rm pl}_{\rm j=0}+\vec p^{\rm pl}_{\rm j\neq0},
\end{equation}
where 
\begin{eqnarray}
	&\displaystyle
	\vec p^{\rm pl}_{\rm j=0}={\vec p^{\rm sph,out}_{\rm eff}}=\vec e_z\varepsilon_0\varepsilon_e\alpha^{\rm pl}_{\rm j=0}E_0\e^{-\i\omega t},\\	
	&\displaystyle
	\vec p^{\rm pl}_{\rm j\neq0}={\vec p^{\rm sph,in}_{\rm eff}}=\vec e_z\varepsilon_0\varepsilon_e\alpha^{\rm pl}_{\rm j\neq0}E_0\e^{-\i\omega t}
\end{eqnarray} 
are the electric dipole moments of the current-free (electrodynamically incompliant) and current-sourced (electrodynamically compliant) scattered fields given in terms of their polarizabilities
\begin{eqnarray}
	&\displaystyle
	\alpha^{\rm pl}_{\rm j=0}=-\frac{2\pi R^3}{\i f_{\rm rad}(R)}=\frac{3\i\pi}{k_e^3},\\	
	&\displaystyle
	\alpha^{\rm pl}_{\rm j\neq0}=-\alpha^{\rm pl}_{\rm j=0}\frac{1+\i A_{\rm qs} f_{\rm rad}(R)}{1-\i A_{\rm qs} f_{\rm rad}(R)}.
\end{eqnarray} 
This separation is further confirmed by Mie theory that provides the exact multipole solution for scattering of electromagnetic waves on a spherical particle. Following it, spherically outgoing parts of the incident fields experience current-free scattering for any polarization and orbital structure \cite{Akimov:2024-2}.   

By interfering, $\vec p^{\rm pl}_{\rm j=0}$ and $\vec p^{\rm pl}_{\rm j\neq0}$ make the resultant electric dipole moment $\vec p^{\rm pl}_{\rm eff}$ resonant and, hence, define all spectral peculiarities in the scattering and absorption cross-sections for deeply subwavelength spheres \cite{Landau:1984,Bohren:1998,Jackson:1999}:
\begin{eqnarray}
	&\displaystyle
	\sigma_{\rm sca}^{\rm pl}=\frac{3\pi}{2k_e^2}\left|\frac{\vec p^{\rm pl}_{\rm eff}\cdot \vec E_0^*}{\varepsilon_0\varepsilon_e\alpha^{\rm pl}_{\rm j=0}|E_0|^2 }\right|^2,\\
	&\displaystyle
	\sigma_{\rm abs}^{\rm pl}=\frac{3\pi}{k_e^2}{\rm Re}\left(\frac{\vec p^{\rm pl}_{\rm eff}\cdot \vec E_0^*}{\varepsilon_0\varepsilon_e\alpha^{\rm pl}_{\rm j=0}|E_0|^2 }\right)-\sigma_{\rm sca}^{\rm pl}.
\end{eqnarray} 
For instance, the destructive interference of the two moments, $\vec p^{\rm pl}_{\rm j\neq0}=-\vec p^{\rm pl}_{\rm j=0}$, results in the nonradiating states of a purely electric dipole structure that do not require toroidal or any other higher-order moments. These states exist at either $k_e=0$ or $\varepsilon_i=\varepsilon_e$. The constructive interference of the two moments, $\vec p^{\rm pl}_{\rm j\neq0}=\vec p^{\rm pl}_{\rm j=0}$, results in the super-radiating state that requires $\varepsilon_i=-2\varepsilon_e$.   

Noteworthy that amongst the two electric dipole moments only the current-sourced one, $\vec p^{\rm pl}_{\rm j\neq0}$, depends on the sphere permittivity and, thus, solely defines its effect on the scattering and absorption cross-sections.
As $A_{\rm qs}$ is resonant to $\varepsilon_i$ given by the pole $(\varepsilon_i+ 2\varepsilon_e)^{-1}$ in Eq.~(\ref{AB_qs}), the effect of $\varepsilon_i$ on the cross-sections differs inside the resonance and out of it. 

\begin{figure}[t]
	\centering\includegraphics[width=0.46\textwidth,clip,trim={1.5cm 1.1cm 3.9cm 0cm}]{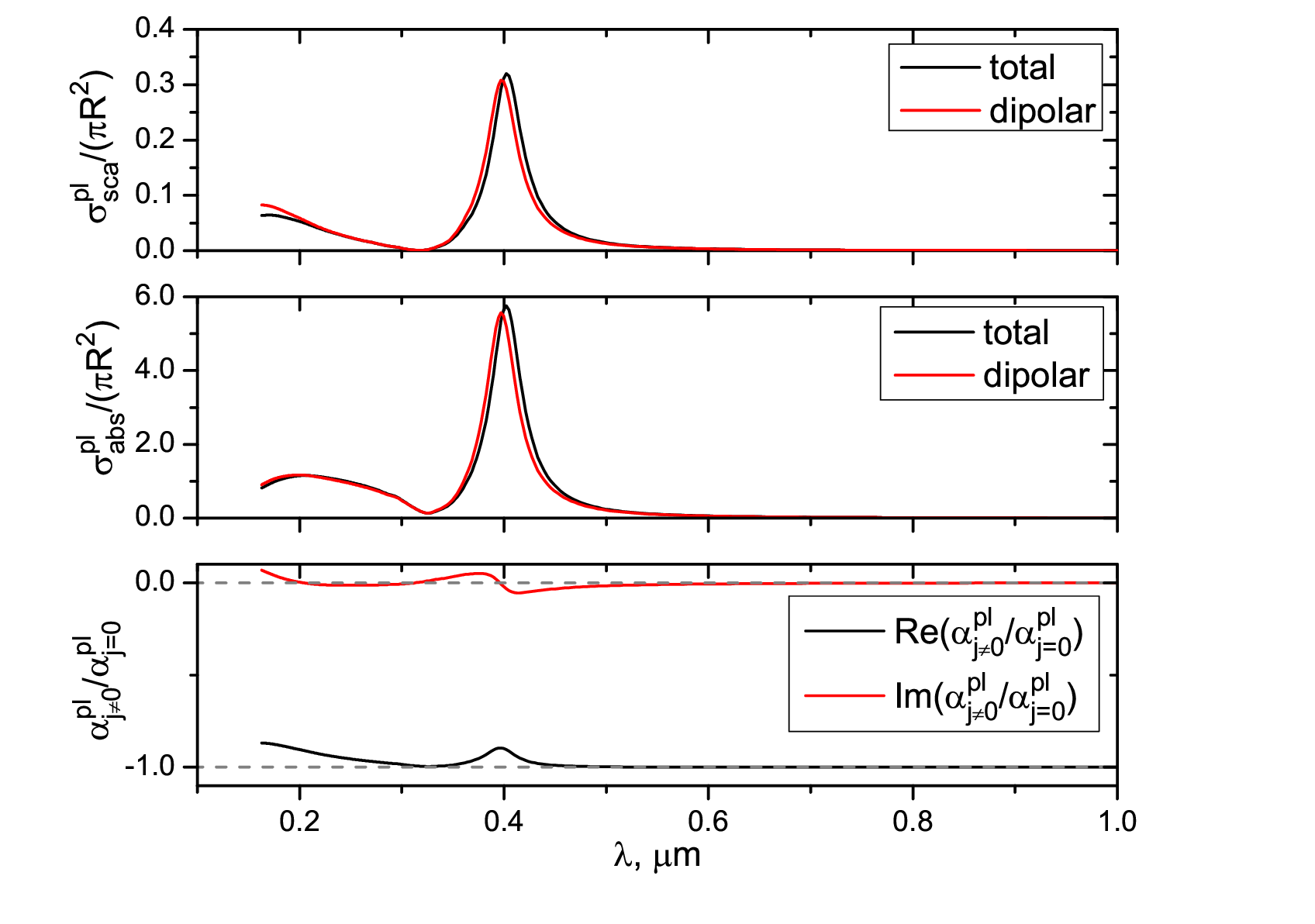}
	\caption{Scattering and absorption cross-sections of silver nanoparticles of $R=10$ nm embedded in silicon dioxide as functions of the wavelength of incident plane waves. Total cross-sections are shown for comparison and calculated based on the Mie theory that includes all types of multipole fields. The dipolar contributions are calculated based on the electric dipole polarizabilities $\alpha^{\rm pl}_{\rm j=0}$ and $\alpha^{\rm pl}_{\rm j\neq0}$. 
	\label{Fig2}}
\end{figure}

For large wavelength out of the resonance, when $|A_{\rm qs}|\sim 1$, the current-sourced electric dipole moment $\vec p^{\rm pl}_{\rm j\neq0}\approx-\vec p^{\rm pl}_{\rm j=0}$. This regime, known as the Rayleigh scattering, is influenced by the nonradiating state with $k_e=0$. The above state is material-independent and common to all subwavelength particles. This regime features $\sigma_{\rm abs}^{\rm pl},\sigma_{\rm abs}^{\rm pl}\ll\pi R^2$ and can be seen in Fig.~\ref{Fig2} for $R=10$ nm Ag nanoparticles embedded in SiO$_2$ at $\lambda>400$ nm when $|\varepsilon_i|>2\varepsilon_e$. 

In the resonant wavelength range, where $|A_{\rm qs}|>1$, the current-sourced electric dipole moment deviates more from its destructive value $-\vec p^{\rm pl}_{\rm j=0}$, giving rise to stronger $\vec p^{\rm pl}_{\rm eff}$ and, hence, scattering and absorption. This regime is influenced by the super-radiating state and seen in Fig.~\ref{Fig2} around the surface-plasmon-polariton resonance at $\lambda=$395 nm with the maximum scattering and absorption cross-sections of $0.3\pi R^2$ and $5.6\pi R^2$ when $\varepsilon_i\approx -2\varepsilon_e$.

For smaller wavelength out of the resonance, at which $|A_{\rm qs}|$ drops below 1, the destructive interference restores influenced by the nonradiating state with $\varepsilon_i=\varepsilon_e$. The corresponding anapole scattering features strong suppression of $\sigma_{\rm sca}^{\rm pl}$ and $\sigma_{\rm abs}^{\rm pl}$, as seen in Fig.~\ref{Fig2} at $\lambda\approx$ 318 nm when $\varepsilon_i\approx\varepsilon_e$. Noteworthy that the quadratic dependence of $\sigma_{\rm sca}^{\rm pl}$ on $\vec p_{\rm eff}^{\rm pl}$ makes its drop deeper compared to $\sigma_{\rm abs}^{\rm pl}$ that features dominant linear dependence on $\vec p_{\rm eff}^{\rm pl}$. For instance, the anapole scattering shown in Fig.~\ref{Fig2} experiences $\sim 500$ times reduction for $\sigma_{\rm sca}^{\rm pl}$ and only $\sim 40$ times for $\sigma_{\rm abs}^{\rm pl}$ with respect to their resonant values. 

\section{Conclusion}

In conclusion, classical scattering problems being the model tasks that use free electromagnetic fields for incident radiation inevitably face the issue of electrodynamically incompliant free scattered fields. 
This issue makes the conventional electrodynamic current moment analysis generally inapplicable to the scattered fields.
As a result, the independent current-sourced and current-free field moments should be used instead. 
Together, the two moments completely define all the spectral features, as has been demonstrated in the paper for the classical case of dipole scattering by deeply subwavelength spherical particles.   

\appendix

\section{Exact solution for induced fields}\label{AppA}

Exact solution for the electromagnetic fields induced by free incident fields of arbitrary structure can be obtained from Maxwell's equations (\ref{Maxwell}).
Inside domains with uniform dielectric permittivity $\varepsilon$, harmonic $\vec E$ and $\vec H$ can be written in terms of the transverse magnetic (TM) and transverse electric (TE) fields as follows \cite{Akimov:2024-2,Akimov:2014}: 
\begin{eqnarray}
	&\displaystyle 
	\vec H=\left(\vec H^{\rm TM}-\i\frac{\lambda}{2\pi} \sqrt{\frac{\varepsilon_0}{\mu_0}}\nabla\times \vec E^{\rm TE}\right)\e^{-\i\omega t},\label{H}\\
	&\displaystyle 
	\vec E=\left(\vec E^{\rm TE}+\i\frac{\lambda}{2\pi\varepsilon}\sqrt{\frac{\mu_0}{\varepsilon_0}}\nabla\times \vec H^{\rm TM}\right)\e^{-\i\omega t},\label{E}
\end{eqnarray}
where $\vec H^{\rm TM}$ and $\vec E^{\rm TE}$ are the amplitudes of the governing fields for the TM and TE polarizations.
In the spherical geometry, the governing TM and TE fields can be decomposed over the vector spherical harmonics of different orbital and azimuthal indices $l, m$ \cite{Akimov:2024-2,Akimov:2014}:
\begin{eqnarray}
	&\displaystyle 
	\vec H^{\rm TM}=\sum_{l=0}^\infty\sum_{m=-l}^l\sum_{j=1}^2
{H}_{lm}^{(j)}h_l^{(j)}(k r)\vec Y_{lm}^{(3)}(\theta,\phi),\label{H_TM_lm}\\
	&\displaystyle 
	\vec E^{\rm TE}=\sum_{l=0}^\infty\sum_{m=-l}^l\sum_{j=1}^2
{E}_{lm}^{(j)}h_l^{(j)}(k r)\vec Y_{lm}^{(3)}(\theta,\phi).\label{E_TE_lm}
\end{eqnarray}
In this decomposition, the angular dependence of the excited fields is fully given by the vector spherical harmonics \cite{Barrera:1985} 
\begin{equation}
\vec Y_{lm}^{(3)}(\theta,\phi)=\vec e_\phi\frac{\pard Y_{lm}(\theta,\phi)}{\pard \theta}-\frac{\vec e_\theta}{\sin \theta}\frac{\pard Y_{lm}(\theta,\phi)}{\pard \phi},
\end{equation} 
defined with 
\begin{equation}
	Y_{lm}(\theta,\phi)=\sqrt{\frac{2l+1}{4\pi}\frac{(l-m)!}{(l+m)!}}\,
	P_l^m (\cos\theta)\e^{\i m\phi},
\end{equation}
where $P_l^m (\cos\theta)$ are the associated Legendre polynomials, while the radial dependence is fully given by 
the spherical Hankel functions $h_l^{(1,2)}(k r)$ defined with $k=2\pi\sqrt{\varepsilon}/\lambda$.
The amplitudes ${H}_{lm}^{(1,2)}$ and ${E}_{lm}^{(1,2)}$ are generally independent of each other and give all possible distributions of electromagnetic fields supported by Maxwell's equations for uniform domains. 

Inside the external domain $r>R$, incident fields can possess any values of $[{H}_{lm}^{(1,2)}]_{\rm inc}^{}$ and $[{E}_{lm}^{(1,2)}]_{\rm inc}^{}$ depending on their spatial distribution (for plane waves $[{H}_{lm}^{(1)}]_{\rm inc}^{}$=$[{H}_{lm}^{(2)}]_{\rm inc}^{}$ and $[{E}_{lm}^{(1)}]_{\rm inc}^{}$=$[{E}_{lm}^{(2)}]_{\rm inc}^{}$). Contrary, scattered fields, being spherically diverging, always feature 
$[{H}_{lm}^{(2)}]_{\rm sca}^{}=0$ and $[{E}_{lm}^{(2)}]_{\rm sca}^{}=0$, 
while $[{H}_{lm}^{(1)}]_{\rm sca}^{}$ and $[{E}_{lm}^{(1)}]_{\rm sca}^{}$ appear linear to the incident field amplitudes:
\begin{eqnarray}
	&\displaystyle 
	[{H}_{lm}^{(1)}]_{\rm sca}^{}=-\sum_{j=1}^2 a^{(j)}_{lm}[{H}_{lm}^{(j)}]_{\rm inc}^{},\\
	&\displaystyle 
	[{E}_{lm}^{(1)}]_{\rm sca}^{}=-\sum_{j=1}^2 b^{(j)}_{lm}[{E}_{lm}^{(j)}]_{\rm inc}^{}, 
\end{eqnarray}
where the coefficients $a^{(1,2)}_{lm}$ and $b^{(1,2)}_{lm}$ give contributions of the TM and TE incident fields with different orbital and azimuthal structure to the scattered fields.  

As for the internal domain $r<R$, the fields there always feature $[{H}_{lm}^{(2)}]_{\rm int}^{}=[{H}_{lm}^{(1)}]_{\rm int}^{}$ and $[{E}_{lm}^{(2)}]_{\rm int}^{}=[{E}_{lm}^{(1)}]_{\rm int}^{}$ for finiteness in the sphere's center, $r=0$. Similarly to scattered fields, internal fields appear linear to $[{H}_{lm}^{(1,2)}]_{\rm inc}^{}$ and $[{E}_{lm}^{(1,2)}]_{\rm inc}^{}$:
\begin{eqnarray}
	&\displaystyle 
	[{H}_{lm}^{(1)}]_{\rm int}^{}=\sum_{j=1}^2 d^{(j)}_{lm}[{H}_{lm}^{(j)}]_{\rm inc}^{},\\
	&\displaystyle 
	[{E}_{lm}^{(1)}]_{\rm int}^{}=\sum_{j=1}^2 c^{(j)}_{lm}[{E}_{lm}^{(j)}]_{\rm inc}^{}, 
\end{eqnarray}
where the coefficients $d^{(1,2)}_{lm}$ and $c^{(1,2)}_{lm}$ give the TM and TE contributions of the incident fields with different orbital and azimuthal structure to the internal fields. 

Following the continuity of the tangential electric and magnetic fields on the particle surface $r=R$, the field coefficients given for spherically outgoing incident fields are 
\begin{eqnarray}
	&\displaystyle 
	a^{(1)}_{lm}=b^{(1)}_{lm}=1, \quad c^{(1)}_{lm}=d^{(1)}_{lm}=0. 
\end{eqnarray}
These coefficients correspond to the current-free scattered fields that appear as a part of the trivial solution of Maxwell's equations for all polarizations, orbital and azimuthal compositions. As for spherically incoming incident fields, their contributions are size- and material-dependent following nontrivial solutions of Maxwell's equations with nonzero polarization currents induced in the spherical particle:  
\begin{eqnarray}
	&\displaystyle 
	a_{lm}^{(2)}=\frac{q_i\psi_l(q_i)\zeta_l'(q_e)-q_e\zeta_l(q_e)\psi_l'(q_i)}
	{q_i\psi_l(q_i)\xi_l'(q_e)-q_e\xi_l(q_e)\psi_l'(q_i)},
\label{S:a_l2}\\
	&\displaystyle 
	b_{lm}^{(2)}=\frac{q_e\psi_l(q_i)\zeta_l'(q_e)-q_i\zeta_l(q_e)\psi_l'(q_i)}
	{q_e\psi_l(q_i)\xi_l'(q_e)-q_i\xi_l(q_e)\psi_l'(q_i)},
\label{S:b_l2}\\
	&\displaystyle 
	c_{lm}^{(2)}=\frac{1}{2}\frac{q_i\psi_l(q_e)\zeta_l'(q_e)-q_i\zeta_l(q_e)\psi_l'(q_e)}
	{q_e\psi_l(q_i)\xi_l'(q_e)-q_i\xi_l(q_e)\psi_l'(q_i)},
\label{S:c_l2}\\
	&\displaystyle 
	d_{lm}^{(2)}=\frac{1}{2}\frac{q_i\zeta_l(q_e)\xi_l'(q_e)-q_i\xi_l(q_e)\zeta_l'(q_e)}
	{q_i\psi_l(q_i)\xi_l'(q_e)-q_e\xi_l(q_e)\psi_l'(q_i)},
\label{S:d_l2}
\end{eqnarray}
where $\xi_l(q)=qh_l^{(1)}(q)$, $\zeta_l(q)=qh_l^{(2)}(q)$, $\psi_l(q)=[\xi_l(q)+\zeta_l(q)]/2$,  are the Riccati--Bessel functions, with $q_i=k_i R$ and $q_e=k_e R$.

\section{Dipolar fields}\label{AppB}

In exact solution of Maxwell's equations given by Eqs.~(\ref{H}) and (\ref{E}), the fields possessing a $z$-polarized electric dipole moment are given by the TM contributions with the orbital index $l=1$ and azimuthal index $m=0$:
\begin{eqnarray}
	&\displaystyle 
	\vec H^{\rm TM}_{\rm ED}=\sum_{j=1}^2
{H}_{10}^{(j)}h_1^{(j)}(k r)\vec Y_{10}^{(3)}(\theta,\phi).
\end{eqnarray}
Eventually, spatial distribution of electric field is given by
\begin{eqnarray}
	\vec E_{\rm ED}=\i\frac{\lambda}{2\pi\varepsilon}\sqrt{\frac{\mu_0}{\varepsilon_0}}\nabla\times \vec H^{\rm TM}_{\rm ED}\e^{-\i\omega t} \label{E1}.
\end{eqnarray}

Applying Eq.~(\ref{E1}) to the internal area of the sphere and assuming $|k_i| r\ll1$, we get the field distribution used in Eq.~(\ref{Eint}) for the internal field excited inside the deeply subwavelength particles:
\begin{eqnarray}
	&\displaystyle	
	\vec E_{\rm int}=C_{\rm int}
	(\vec e_r \cos\theta-\vec e_\theta \sin\theta)\e^{-\i\omega t}+O(|k_i r|^2)~~
\end{eqnarray}
with the field amplitude defined as
\begin{equation}
C_{\rm int}=-\i \left(\frac{4}{3\pi}\frac{\mu_0}{\varepsilon_0\varepsilon_i}\right)^{1/2}{H}_{10}^{(1)}.
\end{equation}
For the external area of the sphere, we separately consider spherically outgoing and incoming fields. 
Under $k_e r\ll1$, Eq.~(\ref{E1}) gives us the following distribution for the outgoing dipolar field
\begin{eqnarray}
	&\displaystyle	
	\vec E_{\rm ext}^{\rm sph,out}=-\i\frac{C_{\rm ext}^{\rm sph,out}}{2f_{\rm rad}(r)}	
	[\vec e_r(2+k_e^2 r^2+\i f_{\rm rad}(r))\cos\theta+~~\nonumber\\
	&\displaystyle	
	\vec e_\theta\left(1-\frac{k_e^2 r^2}{2}-\i f_{\rm rad}(r)\right)\sin\theta]\e^{-\i\omega t}+
	O[(k_e r)^\frac{1}{2}]~~~
\end{eqnarray}
with the amplitude of 
\begin{equation}
C_{\rm ext}^{\rm sph,out}=-\i\left(\frac{4}{3\pi}\frac{\mu_0}{\varepsilon_0\varepsilon_e}\right)^{1/2}{{H}_{10}^{(1)}}.
\end{equation}
This field was used in Eq.~(\ref{Esca}) for the scattered field and in Eq.~(\ref{spherical out}) for the spherically outgoing incident field, where we left only the leading real and imaginary terms.
As for spherically incoming external fields under $k_e r\ll1$, Eq.~(\ref{E1}) results in
\begin{eqnarray}
	&\displaystyle	
	\vec E_{\rm ext}^{\rm sph,in}=\i\frac{C_{\rm ext}^{\rm sph,in}}{2f_{\rm rad}(r)}	
	[\vec e_r(2+k_e^2 r^2-\i f_{\rm rad}(r))\cos\theta+~~~\nonumber\\
	&\displaystyle	
	\vec e_\theta\left(1-\frac{k_e^2 r^2}{2}+\i f_{\rm rad}(r)\right)\sin\theta]\e^{-\i\omega t}+
	O[(k_e r)^\frac{1}{2}],~~~~
\end{eqnarray}
where the field amplitude $C_{\rm ext}^{\rm sph,in}$ is given by
\begin{equation}
C_{\rm ext}^{\rm sph,in}=-\i \left(\frac{4}{3\pi}\frac{\mu_0}{\varepsilon_0\varepsilon_e}\right)^{1/2}{{H}_{10}^{(2)}}.
\end{equation}
This distribution was used in Eq.~(\ref{spherical in}) for the spherically incoming incident field, where we left the dominant real and imaginary terms. Regarding the plane field used in Eq.~(\ref{plane incident}) for $\vec E^{\rm pl}_{\rm inc}$, it is given by the superposition of spherically incoming $\vec E_{\rm ext}^{\rm sph,in}$ and outgoing $\vec E_{\rm ext}^{\rm sph,out}$ fields taken with equal amplitudes ${H}_{10}^{(1)}$ and ${H}_{10}^{(2)}$:
\begin{eqnarray}
	&\displaystyle	
	\vec E_{\rm ext}^{\rm pl}=C_{\rm ext}^{\rm pl}
	(\vec e_r \cos\theta-\vec e_\theta \sin\theta)\e^{-\i\omega t}+O[(k_e r)^2],~~
\end{eqnarray}
where the resultant field amplitude is 
\begin{equation}
C_{\rm ext}^{\rm pl}=-\i \left(\frac{4}{3\pi}\frac{\mu_0}{\varepsilon_0\varepsilon_e}\right)^{1/2}{H}_{10}^{(1)}.
\end{equation}
Overall, the leading terms left in the electric fields demonstrated good accuracy for scattering and absorption of light under $R/\lambda<0.06$ as was shown in Fig.~\ref{Fig2} for the case of silver particle embedded in silicon dioxide.
 
\section*{Disclosures} The author declares no conflicts of interest.

\section*{Data availability} No data were generated or analyzed in the presented research.

\end{document}